LETTER

# DNN Transfer Learning based Non-linear Feature Extraction for Acoustic Event Classification

Seongkyu MUN[†], Minkyu SHIN[††], Suwon SHON[††], Wooil KIM[*], David K. HAN[††] *Nonmembers, and*
Hanseok KO[†, ††], *Member*

**SUMMARY** Recent acoustic event classification research has focused on training suitable filters to represent acoustic events. However, due to limited availability of target event databases and linearity of conventional filters, there is still room for improving performance. By exploiting the non-linear modeling of deep neural networks (DNNs) and their ability to learn beyond pre-trained environments, this letter proposes a DNN-based feature extraction scheme for the classification of acoustic events. The effectiveness and robustness to noise of the proposed method are demonstrated using a database of indoor surveillance environments.
*key words: Acoustic event classification, Transfer learning, Deep neural network, Acoustic feature*

## 1. Introduction

Acoustic event classification (AEC) is the autonomous recognition of different events via sound. It has recently attracted attention due to the increased variety of new applications and potential uses [1-4]. As pointed out in previous studies [2-3], the acoustic features conventionally used in AEC have been shown to overcome the limitations of features based on the human auditory system, such as Mel-frequency cepstral coefficients (MFCCs) and perceptual linear prediction (PLP). Unlike the fixed filter structure of features based on the human auditory system, the extraction of conventional AEC features focuses on training suitable filters to represent acoustic events. In filter pre-training, non-negative matrix factorization (NMF) [2], non-negative K-SVD (singular value decomposition) [3] and statistical distribution [4] have been used to extract information that can be used to discriminate target events. The features obtained by the aforementioned approaches have shown better AEC performance compared to the human auditory system based features. However there is still room for improvement in two respects. First, AEC has known issues with 'weakly labeled' databases (DBs) and a deficit of target DBs [5] compared to other audio signal applications such as speech recognition, natural language processing, and speaker recognition. These DB issues can lead to the insufficient or non-generalized pre-training of target event filters. Second, the filtering processes of conventional approaches are based on linear combination. In recent audio detection [6] and signal enhancement [7] researches, non-linear modeling has exhibited improved performance compared to linear approaches. Based on the advantages demonstrated by researches mentioned above, it can be inferred that a filtering process which can model the non-linear relationships among frequency bands may improve AEC performance.

To address these issues, this letter proposes to use a deep neural network (DNN), which is trained using transfer learning, as an acoustic event filter. The 'transfer learning' scheme aims to transfer knowledge between the source domain used for pre-training and the target domain of interest. In computer vision, transfer learning overcomes the deficit of target domain training samples by adapting layer parameters that have been pre-trained for other large-scale DBs [8]. The source domain DB is also referred to as the background or development DB, and the size of the source DB is generally larger than that of the target domain. The success of transfer learning in visual object classification (VOC) has been attributed to the effectiveness of transferring the neural network parameters from the source to the target domain. Therefore, this approach may help to pre-train the acoustic event filter more effectively using transferred parameters which have been already trained to extract discriminative acoustic information from large DBs. Moreover, DNN-based filtering can effectively model the non-linear relationships among frequency bands for AEC by using nonlinear activation functions with a larger number of parameters, compared to NMF or the distribution-based filters [2-4].

## 2. Proposed Feature Extraction for AEC

The acoustic event filter training and the AEC system are depicted in Fig. 1. Unlike the conventional method shown in Fig. 1(a), the proposed method presented in Fig. 1(b) has an additional training step in the source domain. The DNN filter is first trained in the source domain then

---

[†] The author is with the Department of Visual Information Processing, Korea University, Seoul, 136-713, Korea.
[††] The authors are with the School of Electrical Engineering, Korea University, Seoul, 136-713, Korea.
[*] The author is with the Dept. of Computer Science and Engineering, Incheon National University, Incheon, Korea

transferred to the target domain and adapted to target domain classes with additional layers. After the training step, the DNN filter is used to extract features in the AEC system. Details of DNN filter training are provided in Fig. 2. In the source domain, the network is composed of three hidden fully connected layers which use a Sigmoid activation function and a single output layer with a SoftMax function. For filter training in the target domain, similar to the transfer learning in VOC [8], the output layer of the pre-trained network is removed and two hidden fully connected layers and a new single output layer are added to enable adaptation. Because the transferred layers have been pre-trained to classify various classes within the source domain, the layer outputs may capture the discriminative features of different sounds [8]. In target domain training, the outputs of the transferred layer are adapted to target domain labels by using them as inputs for training the additional two hidden layers. In summary, the parameters for layers SL#1-3 are first trained in the source domain then transferred to the target domain and fixed. Only the additional adaptation layers (TL#1-2) are trained using the target domain training data.

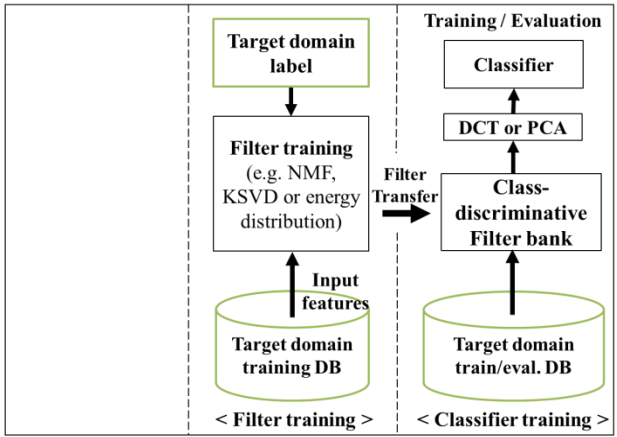
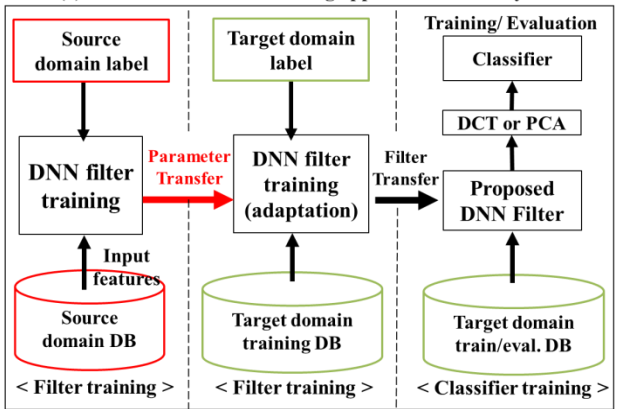

**Fig. 1** Comparison of the conventional and proposed feature extraction methods

After the target domain training step, as depicted in Fig. 3, the output layer and activation functions of the last hidden layer (TL#2) are removed. This process is motivated by bottleneck feature studies [1,9], which follow a similar approach in using DNN mid-layers and demonstrate effective performance. Finally, the five hidden layers from SL#1 to TL#2 are used as a DNN filter and the output values of layer TL#2 without the activation function are used as the input features for the AEC system.

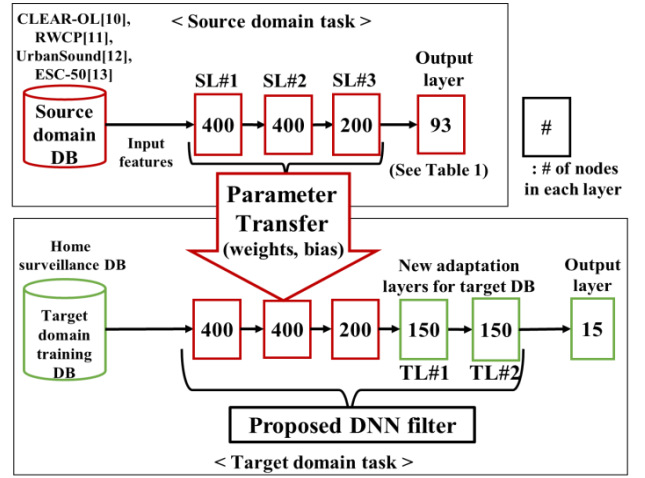

**Fig. 2** The proposed filter training process using a transfer learning based DNN filter

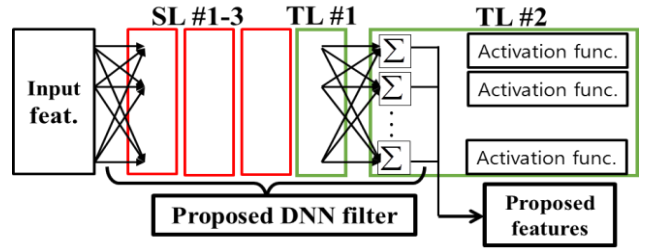

**Fig. 3** The proposed DNN filter and feature extraction after filter training

## 3. Experimental Settings and Results of AEC

For the source domain task, four acoustic DB sets (CLEAR-OL, RWCP, UrbanSound, and ESC-50) were merged [10-13]. Since each DB set has a different wave length for their own classes, the wave length of each class was normalized to about 800 seconds. Classes over 800 seconds were randomly cut into 800 seconds and those under 800 seconds were reproduced by filtering the room impulse response of the office environment (estimated $RT_{60}$ was 0.7sec) as if they were re-recorded in office environments. A total of 93 classes were selected and the wave files were resampled at 16 kHz and 16 bits resolutions. Detail descriptions are shown in Table 1.

For the target domain task, an indoor surveillance DB

was used. The database consists of 15 events (a crying child, breaking glass, water drops, chirping birds, a doorbell, home appliance beeping, screaming, a dog barking, music, speech, a cat meowing, a gunshot, a siren, an explosion, and footsteps). It was collected at various locations by a portable recorder and employed the following datasets: the BBC Sound Effects Library [14], Sound Ideas [15], and the Sony Sound Effects Library [16]. Each event consisted of 150 segments for the training set and 50 segments for the evaluation set (each segment is a wave file of 3 seconds). To evaluate robustness in real life, two noises were chosen from the ETSI background noise DB [17]. From the background noise DB, target class related components were excluded and noises were added to the event DB at 5, 10, 15 dB SNR.

For the source and target domains in DNN filter training, the training procedure periodically evaluated the cross-entropy objective function on a subset of the training set and a validation set (the training and validation set ratio was 4:1 for both the source and target training DBs). The initial learning rates were set to 0.005 and the network was trained until training cross-entropy was stabilized. The learning rates were then divided by 10 and the training procedure was repeated. The momentum parameter and weight decay were set to 0.9 and 0.0005, respectively.

In order to select the input feature type of the DNN filter, we conducted AEC experiments with various types of input, such as magnitudes of discrete Fourier transform (DFT) [2-4,18], real and imaginary values of DFT [19] and waveform [20]. Based on recent AEC research [21], which used multiple frames for input, the evaluation was performed by varying the number of frames for splicing. The structure of DNN filter with output layer (lower part in Fig. 2) was used for the experiment. Details of the feature settings and experimental results are listed in Table 2. Despite conducting experiments with various hyper parameter adjustments, waveform and real /imaginary value inputs did not show better performance compared to the DFT magnitude input. While these raw (or relatively raw) features have potential to improve performance, they have been considered to require a complicated network structure and a very large number of weight parameters (2~18M parameters for waveforms [20], 24M for [19]) for achieving high level of performance. It is noted that the main focus of this letter is on the effectiveness of DNN filter based on transfer learning. Hence, we will conduct additional experiments for various types of input features through a future research.

As mentioned in the previous section, this letter used DNN as feature extractor rather than classifier. This approach differs from the conventional usage of DNN which performs whole process (from input features to output of classification results) by a single network. As an alternative approach to the conventional methods, we believed that the proposed DNN mid-layer information extracting and combining with various types of classifiers and feature transformations could further improve the AEC performance. Therefore, in order to evaluate the performance of the proposed features with additional processes, we used Gaussian mixture model (GMM), support vector machine (SVM) and DNN for classifier comparison. Before classification, discrete cosine transform (DCT) and principal component analysis (PCA) were individually applied to features for redundancy reduction and de-correlation [1-4]. The classification result of each segment is obtained by accumulating the result (probability or SoftMax output of each class) per frame. Details of the classifier settings and experimental results are listed in Table 3.

Table 1. Source domain database description

| DB set | Contents |
|---|---|
| Clear-OL [10] | Alert, cough, door slam, drawer, key, keyboard, knocking, laughing, mouse, page turn, pen drop, phone, printer, speech, switch, clear throat |
| RWCP [11] | Air-cap, bell, break stick, buzzer, castanet, ceramic collision, clap, clock ringing, coin, cymbals, drum, dryer, grinding coffee, kara, maracas, metal collision, article dropping, plastic collision, pump, punch stapler, rubbing, shaver, spray, string, tambourine, toy, whistle, wood collision |
| Urban-Sound [12] | Air-conditioner, dog bark, drilling, engine idling, car horn, jackhammer, children playing, siren, street music, shot |
| ESC-50 [13] | Airplane, breathing, brushing teeth, can opening, cat, chainsaw, chirping birds, church bells, clapping, clock alarm, clock tick, coughing, cow, crackling fire, crickets, crow, door - wood creaks, door knock, drinking – sipping, engine, fireworks, footsteps, frog, hand saw, helicopter, hen, insects (flying), pig, pouring water, rooster, sea waves, sheep, sneezing, snoring, thunderstorm, toilet flush, vacuum cleaner, washing machine, wind |
| **Total 93 classes / The similar classes from the different DB set had been merged / 16 kHz resampled, 16 bit resolution** |||

Table 2 Average acoustic event classification rate [%]
on various input features for the DNN filter with output layer (Fig. 2)

| # of spliced frames for input | **(1)** | (2) | (3) | (1)+(2)+(3) |
|---|---|---|---|---|
| 1 | 88.5 | 60.4 | 64.5 | 68.2 |
| **3** | **89.2** | 60.5 | 65.7 | 69.3 |
| 5 | 88.9 | 61.2 | 65.3 | 69.6 |
| 7 | 88.1 | 61.2 | 66.1 | 71.8 |

(1)DFT mag. [512 points], (2)Waveform [1024]
(3)DFT real & image. [1024=512(real)+512(image.)],
- Input features are normalized based on
the statistics of source and training part of target DB.
- Average accuracy on clean and noisy DB (see details in Table. 4).

Table 3 Average acoustic event classification rate [%] on
combination of various feature transformations and classifiers

| | GMM | **SVM** | DNN |
|---|---|---|---|
| W/o transformation | 89.1 | 89.7 | 89.4 |
| **With DCT** | 94.4 | **95.8** | 95.3 |
| With PCA | 94.0 | 95.1 | 95.4 |

-Average accuracy on clean and noisy DB (see details in Table. 4).
-DNN filter with the best performance setting in Table 2 was used.
-GMM with 512 mixtures (diag. cov.), SVM with radial basis function kernels and DNN classifier of three hidden layers (300-300-100) were empirically selected.
-150 points of DCT is applied and leading 50 points are selected as features. / PCA also reduced feature dimensions to 50.
-The hyper parameters of the classifiers and transformations were tuned using 5-fold cross-validation.

Based on the aforementioned two experimental results, the best performance setting (3-spliced frames, SVM with DCT) was selected and compared with other conventional algorithms. This letter compares the average accuracy for all events for the conventional and proposed methods. Table 4 presents the segment-based classification accuracy. It was found that the proposed method achieved higher accuracy than the other approaches in the clean and all noise conditions. In comparison to the NMF [2] and K-SVD [3] linear filter results, the proposed non-liner filter produced a more noise robust performance. In addition, compared with other DNN-based feature extraction methods, such as the Deep Belief Network (DBN) feature, which is used for music genre classification [18], and the DNN bottleneck feature [1], the proposed method demonstrated improved accuracy by effectively utilizing the information transferred from the source domain.

**Table 4** Average acoustic event classification rate [%] for ETSI background noise using various features with SVM classifier

|  | Living room noise | | | Office noise | | | Clean DB | Average |
|---|---|---|---|---|---|---|---|---|
| SNR [dB] | 5 | 10 | 15 | 5 | 10 | 15 | | |
| MFCC | 79.7 | 85.5 | 94.5 | 81.1 | 87.6 | 95.1 | 96.1 | 88.5 |
| NMF [2] | 85.1 | 87.1 | 94.8 | 89.1 | 92.3 | 96.1 | 98.5 | 91.9 |
| K-SVD [3] | 86.0 | 90.3 | 95.2 | 89.3 | 91.9 | 96.3 | 98.3 | 92.5 |
| DBN feature [18] | 86.4 | 89.9 | 93.9 | 89.9 | 93.3 | 95.7 | 96.4 | 92.2 |
| DNN-bottleneck feature [1] | 86.3 | 90.9 | 95.5 | 90.7 | 92.5 | 95.9 | 96.5 | 92.6 |
| [A] | 81.1 | 87.1 | 94.5 | 85.1 | 87.6 | 94.5 | 94.8 | 89.2 |
| [B] | 86.1 | 91.1 | 95.1 | 91.1 | 95.7 | 95.7 | 96.4 | 93.0 |
| **[C]** | **92.5** | **96.3** | **96.3** | **93.7** | **96.5** | **96.5** | **98.9** | **95.8** |

[A] DNN filter w/o removing activations and output layer
[B] Proposed features without transfer learning
**[C] Proposed features**

## 4. Conclusion

To improve AEC performance, this letter proposed a novel DNN filter training framework employing transfer learning. By utilizing the information transferred from the source domain, the proposed feature extraction was characterized by improved AEC accuracy in indoor surveillance experiments.

Once DNN filter training has been completed in the source domain, this DNN filter can be utilized in other domains, repeatedly. Therefore, future work will investigate an effective transfer learning scheme for various acoustic applications and determine how performance changes depending on the configuration of the data.

## Acknowledgments

This material is based upon work supported by the Air Force Office of Scientific Research under award number FA2386-16-1-4130 and authors would like to thank the anonymous reviewers for their valuable comments.